\documentclass[10pt]{article}
\usepackage{amsfonts,amssymb,epsfig}
\usepackage{amsmath}
\def\sqr#1#2{{\vcenter{\hrule height.#2pt\hbox{\vrule width.#2pt
height#1pt \kern#1pt \vrule width.#2pt}\hrule height.#2pt}}}
\def\square{\mathchoice\sqr64\sqr64\sqr{4.2}3\sqr{3.0}3 \ \!} 

\newtheorem{THEOREM}{Theorem}[section]  
\newenvironment{theorem}{\begin{THEOREM} \hspace{-.85em} {\bf :} 
}%
                        {\end{THEOREM}}
\newtheorem{LEMMA}[THEOREM]{Lemma}
\newenvironment{lemma}{\begin{LEMMA} \hspace{-.85em} {\bf :} }%
                      {\end{LEMMA}}
\newtheorem{COROLLARY}[THEOREM]{Corollary}
\newenvironment{corollary}{\begin{COROLLARY} \hspace{-.85em} {\bf 
:} }%
                          {\end{COROLLARY}}
\newtheorem{PROPOSITION}[THEOREM]{Proposition}
\newenvironment{proposition}{\begin{PROPOSITION} \hspace{-.85em} 
{\bf :} }%
                            {\end{PROPOSITION}}
\newtheorem{DEFINITION}[THEOREM]{Definition}
\newenvironment{definition}{\begin{DEFINITION} \hspace{-.85em} {\bf 
:} \rm}%
                            {\end{DEFINITION}}
\newtheorem{EXAMPLE}[THEOREM]{Example}
\newenvironment{example}{\begin{EXAMPLE} \hspace{-.85em} {\bf :} 
\rm}%
                            {\end{EXAMPLE}}
\newtheorem{CONJECTURE}[THEOREM]{Conjecture}
\newenvironment{conjecture}{\begin{CONJECTURE} \hspace{-.85em} 
{\bf :} \rm}%
                            {\end{CONJECTURE}}
\newtheorem{PROBLEM}[THEOREM]{Problem}
\newenvironment{problem}{\begin{PROBLEM} \hspace{-.85em} {\bf :} 
\rm}%
                            {\end{PROBLEM}}
\newtheorem{REMARK}[THEOREM]{Remark}
\newenvironment{remark}{\begin{REMARK} \hspace{-.85em} {\bf :} 
\rm}%
                            {\end{REMARK}}
\newtheorem{CONCLUSION}[THEOREM]{Conclusion}
\newenvironment{conclusion}{\begin{CONCLUSION} \hspace{-.85em} {\bf :} 
\rm}%
                            {\end{CONCLUSION}}
\newcommand{\thm}{\begin{theorem}}
\newcommand{\lem}{\begin{lemma}}
\newcommand{\pro}{\begin{proposition}}
\newcommand{\dfn}{\begin{definition}}
\newcommand{\rem}{\begin{remark}}
\newcommand{\con}{\begin{conclusion}}
\newcommand{\xam}{\begin{example}}
\newcommand{\cnj}{\begin{conjecture}}
\newcommand{\prb}{\begin{problem}}
\newcommand{\cor}{\begin{corollary}}

\newcommand{\ethm}{\end{theorem}}
\newcommand{\elem}{\end{lemma}}
\newcommand{\epro}{\end{proposition}}
\newcommand{\edfn}{\bbox\end{definition}}
\newcommand{\erem}{\bbox\end{remark}}
\newcommand{\econ}{\bbox\end{conclusion}}
\newcommand{\exam}{\bbox\end{example}}
\newcommand{\ecnj}{\bbox\end{conjecture}}
\newcommand{\eprb}{\bbox\end{problem}}
\newcommand{\ecor}{\end{corollary}}

\newcommand{\beqn}{\begin{equation}}
\newcommand{\eeqn}{\end{equation}}

\newcommand{\bbox}{\vrule height7pt width4pt depth1pt}

\newcommand{\sect}[1]{\setcounter{equation}{0}\bigskip\medskip
\section{#1}\smallskip}

\def\br{\begin{eqnarray}}
\def\er{\end{eqnarray}}
\def\brn{\begin{eqnarray*}}
\def\ern{\end{eqnarray*}}
\def\er{\end{eqnarray}}
\def\beq{\begin{equation}}
\def\eeq{\end{equation}}

\def\vp{\varphi}

\def\a{\alpha}
\def\b{\beta}

\def\e{\boldsymbol {e}}
\def\p{\overline {\psi}}
\def\tx{\tilde {x}}
\def\ty{\tilde {y}}
\def\tz{\tilde {z}}
\def\tt{\tilde {t}}
\def\tbr{\tilde {\mathbf{r}}}
\def\bv{\mathbf{v}}
\def\bbr{\mathbf{r}}
\def\bpsi{\boldsymbol{\psi}}
\def\bR{\boldsymbol{R}}
\def\dbpsi{\boldsymbol{\dot{\psi}}}
\def\ddbpsi{\boldsymbol{\ddot{\psi}}}
\def\dddbpsi{\boldsymbol{\dot{\ddot{\psi}}}}

\def\dps{\dot{\overline {\psi}}}

\title{{\bf On the derivation of the equations of motion in  
theories of gravity.}\\
}
\author{
\small {\sf Shmuel Kaniel \thanks {kaniel@math.huji.ac.il}
 and Yakov Itin \thanks {itin@math.huji.ac.il}
}\\
\small {\sf Institute of Mathematics}\\
\small {\sf Hebrew University of Jerusalem}\\
\small {\sf Givat Ram, Jerusalem 91904, Israel}\\
}
\begin{document}  

\newcommand{\bi}[1]{\bibitem{#1}}
\date{\today}

\maketitle
\begin{abstract}
General Relativity is unique, among the class of field theories, in 
the treatment of the equations of motion. 
The equations of motion 
of massive particles are completely determined by the field equation.
Einstein's field equations, as well as most field equations in gravity 
theory, have a specific analytical form: 
They are linear in the second 
order derivatives and quadratic in the first order, with coefficients 
that depend on the variables. 
We utilize this particular form and 
propose  for the $N$-body problem  of the equations that are Lorentz invariant 
a novel algorithm for the derivation of the equations of 
motion from the field equations. 
It is:
 \begin{itemize}
\item[\bf{1.}] Compute a static, spherically symmetric solution 
of the field equation. 
It will be singular at the origin. 
This will be taken to be the field generated by a single particle.
\item[\bf{2.}] Move the solution on a trajectory 
$ {\psi(t)}$ and apply the instantaneous Lorentz 
transformation based on instantaneous velocity $\dot{\psi}(t)$.
\item[\bf{3.}] Take, as first approximation, the field generated 
by $N$ particles to be the superposition of the fields generated 
by the  single particles. 
\item[\bf{4.}] Compute the leading part of the equation. 
Hopefully, only terms that involves $\ddot{\psi}(t)$ will be dominant. 
This is the ``inertial'' part.
\item[\bf{5.}]Compute  by the quadratic part of the equation.
This is the agent of  the ``force''.
\item[\bf{6.}] Equate for each singularity, the highest order terms 
of the singularities that came from the linear part and the 
quadratic parts, respectively. 
This is an equation between the inertial part and the force.  
\end{itemize}
The algorithm was applied to 
Einstein equations. 
The approximate evolution of scalar curvature  
lends, in turn, to an invariant scalar equation.  
The algorithm for it  did produce Newton's law of gravitation. 
This is, also,  the starting point for the embedding the trajectories in a 
common field. 
\end{abstract}
\sect{Introduction}          
General Relativity (GR) is unique among the class of field theories 
in the treatment of the equations of motion. 
The equations of motion of massive particles are completely 
determined by the field equation. \\
By comparison, classical electrodynamics postulates, in addition 
to the Maxwell field equations,  the force 
(Coulomb or Lorentz) as well the form of the inertial term. \\
The derivation of the equation of motion for massive particles 
from the GR field equations was obtained  in the pioneering work 
of Einstein, Infeld and Hoffman \cite{EIH1} of 1939. 
The  particles were treated as  singular points of the metric field. 
The first order approximation of the field equations 
resulted in the verification of  Newton law of attraction 
between the singular points.
In subsequent papers \cite{EIH2},\cite{EIH3} the higher order 
approximations to the geodesic line equation (the equations of motion) 
was calculated. 
The successive approximations were all derived without 
additional hypothesis (i.e. the Geodesic Postulate).
The method used in the cited papers  is known as the {\it EIH-procedure}.  
The derivation is rather formal.   
A similar approach was adopted by  Fock \cite{Fo}. 
He, however, studied Einstein field equations with the RHS-terms 
- the tensor of energy-momentum  of matter. 
Sternberg \cite{St} proved the geodesic postulate in a general 
form for a system that 
incorporates also Yang-Mills fields. 
He refers to the establishment of the postulate as to the EIH condition. 
In his work there is no explicit computation of the trajectories for 
the $N$-body problem. 
The length and complexity of EIH analysis undoubtedly discouraged 
further research this area.\\
The derivability of the equations of motion is attributed to two factors:
\begin{itemize}
\item[(i)] The GR field equations are non-linear (as opposed to 
the other classical field equations). 
As a result, the linear combination of 1-point static 
solutions does not provide a new static solution. 
\item[(ii)] The system of the GR field equations is   subject 
to additional  identities (Bianchi) which reduce the number of 
independent equations.  
The conservation laws (which rest upon these identities) play a 
crucial role in the original   EIH-procedure.
\end{itemize}
Thus two  natural questions arise:
\begin{itemize}
\item[(i)] What other non-linear field equations have the property of 
derivability of the equations of motion?
\item[(ii)] Is there a way to embed the moving singularities 
in a field satisfying the prescribed field equation.
\end{itemize}
In this paper we deal with the first question and touch upon the second one. 
First we define a suitable class of the field equations. 
These are linear in the second order derivatives and quadratic in 
the first order derivatives both with  coefficients that depends 
only on the field variables. 
It is also required that the equations are Lorentz invariant. 
It will be demonstrated in the sequel that  the linear part is 
{\it the agent of  inertia} and the quadratic part is 
{\it the agent of the  force}. 
Let the field equation have a static spherical symmetric 
solution (exact or approximate) which is singular at a point.
 Accordingly, by the Einstein description,  this is the field 
of one  particle. 
The field of a particle moving with constant velocity is taken to 
be the Lorentz transform of the static field. 
If it moves on a curved trajectory $\bpsi(t)$ then the Lorentz transformation 
with the instantaneous velocity $\dbpsi$ is taken. 
The field of $N$ particles is taken to be, approximately, 
the superposition of  the fields of the single particles. 
When this  is inserted to the field equation two singular 
expressions emerge: one from the linear part and the other 
one from the quadratic part. 
By equating the highest order terms of these singularities  
the strength of the singularity of the solution is reduced. 
It turns out that this balance equation is  Newton's law of attraction.
\sect{The general form of field equations and of their solutions.}       
We will require the following for  the field equation:
\begin{itemize}
\item [{\bf(i)}] PDE of the second order,
\item [{\bf(ii)}] Linear in the second order derivatives with coefficients 
that depend only on the field variables,
\item [{\bf(iii)}] Quadratic in the first order derivatives.
\end{itemize}
The last condition can be motivated by  the following dimensional 
argument.\footnote{It was communicated to us by F.\,W.~ Hehl.}  
Let the field variable $\Phi$ will be dimensionless. 
Thus, the second order derivatives $\Phi_{,\mu\nu}$ have the 
dimension of $[l^{-2}]$. 
The only polynomial of the first derivatives that has  the same 
dimension is quadratic in the first order derivatives:  
$\Phi_{,\mu}\Phi_{,\nu}$. 
Thus, in order to have a field equation with dimensionless 
constants we have to consider a PDE, which is linear in the 
second order derivatives and quadratic in the first order derivatives.\\
As a consequence we will restrict ourselves to  field  
equations of the form:
\begin{equation}\label{fe2} 
a\Phi_{,\mu\nu}-b\Phi_{,\mu}\Phi_{,\nu}=0,
\end{equation}
where the coefficients $a$ and $b$ are the dimensionless functions 
of the field variable (or constants)
$$a=a(\Phi),\qquad b=b(\Phi).$$
The covariance condition as well as the existence of an 
action functional provide certain restrictions on the 
coefficient  $a(\Phi)$ and $b(\Phi)$. 
Note that the form (\ref{fe2}) is still symbolic, since 
the field $\Phi$ as well as the coefficient functions $a$ and 
$b$ can possess the interior or coordinate indices and 
these indices should be contracted in a proper manner.\\
Note also  that the Einstein equation does have the form (\ref{fe2}).\\ 
Let us turn now to the properties of the possible physical relevant 
solutions of the field equation (\ref{fe2}).\\
We will apply {\it the Einstein model of a particle}: it is a smooth 
solution of the field equation which tends to a constant at 
infinity and singular at a point. 
In this model the field pertaining to one particle has only one singularity. 
In the proper coordinates, the field around the singularity 
is spherically symmetric and tends to infinity at the singular point. 
The field pertaining to $N$ particles is approximated by the 
superposition of rigid motions of of the fields pertaining to all 
the particles. 
The trajectory of a singularity, which is taken to 
be the trajectory of the correspondent particle, is affected 
by the presence of the fields pertaining to the other particles. 
This means that the field equation should be non-linear 
(for a linear equation the fields can not affect each other).\\ 
Let us also require that a static spherical symmetric solution 
of (\ref{fe2})  has a long distance expansion 
\begin{equation}\label{expan} 
\Phi=\Phi_0+\frac {A_1}{r}+\frac {A_2}{r^2}+\cdots + 
\frac {A_n}{r^n} + \cdots,
\end{equation}
where $\{A_n,n=1,2,...\}$ are constants.\\
It is enough to require the convergence of this expansion for 
distances greater of some characteristic length of the system.
The  dimension of $A_n$ is  $[l^n]$. 
A second order equation usually produces  two independent 
constants of integration. 
Thus  the set of constants can be normalized:
\begin{equation}\label{norm} 
 A_1=m, \qquad A_n=C_n m^n,\quad n=2,3,...,
\end{equation}
where the dimension of the constant $m$ is equal to $[l]$ and
the constants $C_i$ are dimensionless.\\
Einstein derives the equation of motion by considering the 
integral relations which are implied by the conservation laws. 
We propose a different procedure to be exhibited in the sequel. 
\sect{A Qualitative Description of the Algorithm}
In this section we describe a novel algorithm for deriving 
the equations of motion from the field equation. 
We consider the general field equation (\ref{fe2}) 
without specification of the tensorial nature of the field variable $\Phi$. 
We do require, however, that (\ref{fe2}) is Lorentz invariant. \\
Let the field equation (\ref{fe2}) have a static spherical symmetric solution 
$\Phi(\bbr-\bbr_0)$ with a singularity located at $\bbr=\bbr_0$.  
Denote the Lorentz transformation based on the velocity $v$ by $L_v$.  
Consequently, if  $\Phi=\Phi(\bbr-\bbr_0)$ is a time independent solution 
of  (\ref{fe2}) then $ L_v\Phi$ is also a solution of the same equation 
for an arbitrary Lorentz transformation $L_v$. 
(If $\Phi$ is multi-component like a tensor then $L_v$ involves 
not only a coordinate change but, also, a transformation of 
components of $\Phi$.) 
The solution $ L_v\Phi$  describes the field of a pointwise 
singularity moving with a constant velocity $v$ on the trajectory 
$\bpsi=\bv t$. 
Let us try to construct a generalization of Lorentz transformation 
where the origin moves on a general trajectory $\bpsi=\bpsi(t)$. 
Denote such a transformation by  $N_\psi$. The choice 
$$N_\psi=L_{\dot\psi}$$ 
is a plausible candidate. 
$N_\psi\Phi$ is a rigid motion of the field.\\
Now substitute $N_\psi\Phi$ in (\ref{fe2}). 
If $\dot \psi=const$ then $N_\psi\Phi$ is also a solution of (\ref{fe2}). 
If not, then the linear part produces extra terms. 
These extra terms come from two sources:
\begin{itemize}
\item [{\bf(i)}] {\it The derivatives of the Lorentz root} 
$\sqrt{1-{\dot\psi}^2}$. 
In our consideration the velocity of the particle $\dot\psi$ as well as 
its time and spatial derivatives are assumed to be small.  
It follows that the derivatives of the root are of the form 
${\cal O}({\ddot{\psi}} \dot\psi)$. Thus they may be discarded.
\item [{\bf(ii)}] {\it The linear  part.}
Since $\Phi$ is time independent the first derivatives of 
$\Phi$ are only the spatial ones. 
Thus, the first derivatives of of $L_{\dot\psi}\Phi(x)$ 
involve spatial derivatives of $\Phi$ multiplied by $\dot\psi$. 
The second derivatives of $L_{\dot\psi}\Phi(x)$ cancel each other 
the same way as in a Lorentz transformed solution $L_v\Phi(x)$ 
with one exception: 
The second derivative $\ddot\psi$ multiplied by spatial 
first derivatives of $\Phi$ do remain. 
This extra term that come from the linear part is the {\it agent of inertia}.
\item [{\bf(iii)}] {\it The quadratic part.} 
It involves only first order derivatives. 
Consequently the fact that $ \dot\psi$ is variable does not affect it's form.
\end{itemize}
Construct now  a solution that describes the field of $N$ particles, 
i.e. a field with $N$ singular points. It can be approximated by a 
superposition of 1-singular solutions moving on arbitrary trajectories 
${}^{(j)}\psi={}^{(j)}\psi(t)$
$$\Phi=\sum_j L_{{}^{(j)}\dot\psi}{}^{(j)}\Phi(x).$$
Substituting this approximate solution in (\ref{fe2}) we obtain, in 
the linear part, only the second derivatives ${}^{(j)}\ddot\psi$ 
multiplied by the spatial first derivatives of $\Phi$. \\
Consider the quadratic part. 
It is composed of the first order derivatives of 
${}^{(j)}\Phi(x)$ multiplied by the first order derivatives 
of $L_{{}^{(k)}\Phi}{}^{(k)}\Phi(x)$. If $j=k$, since  ${}^{(k)}\Phi(x)$ is 
a solution of  (\ref{fe2}), these products are cancelled by the 
linear part operating   on ${}^{(k)}\Phi(x)$. 
If $j\ne k$ the products will, hopefully, be an approximation 
to the interaction between the $j$-th and the $k$-th particles.\\
Near the $k$-th singularity, for the linear part, 
 only the terms coming from   ${}^{(k)}\Phi(x)$ will be dominant. 
Likewise, for the quadratic part, only the terms involving the 
derivatives of ${}^{(k)}\Phi(x)$ will be dominant. 
Equating, near the singularity, of the two terms above 
(again to the leading order) should, hopefully, result 
in the Newtonian law of attraction.\\ 
Let us summarize the  novel algorithm for the derivation 
of equations of motion.
\begin{itemize}
\item[\bf{1}] Compute a static, spherically symmetric 
solution of the field equation. 
It will be singular at the origin. 
This will be taken to be the field generated by a single particle.
\item[\bf{2}] Move the solution on a trajectory $ \bpsi(t)$ 
and apply the instantaneous Lorentz transformation based on $\dbpsi(t)$.
\item[\bf{3}] Take the field generated by $n$ particles 
to be the superposition of the fields generated by the  single particles. 
\item[\bf{4}] Compute the leading part of the equation. 
Hopefully, only terms that involves $ \ddbpsi$ will be dominant.
\item[\bf{5}] Compute the ``force'' between the particles by 
the quadratic part of the equation.
\item[\bf{6}] Equate for each singularity, the highest 
order terms of the singularities that came from the linear 
part and the quadratic parts, respectively. This is an 
equation between the inertial part and the force. 
\end{itemize}
\sect{Einsteinian gravity}                            
Let us apply the procedure described above to the Einsteinian gravity.
Our first step is  to construct an approximated 
dynamical $N$-particle solution from the static 1-particle solution.  
The Einstein field equation in vacuo is 
\begin{equation}\label{3-1}
R_{ik}=0.
\end{equation}
Consider the Schwarzschild solution in isotropic coordinates
\begin{equation}\label{3-2}
ds^2=\Big(\frac{1-m/4r}{1+m/4r}\Big)^2dt^2-(1+m/4r)^4(dx^2+dy^2+dz^2).
\end{equation}
We will use the general diagonal metric of the form
\begin{equation}\label{3-3}
ds^2=e^{2f}dt^2-e^{2g}(dx^2+dy^2+dz^2).
\end{equation}
The static solution (\ref{3-2}) has the long-distance expansion 
\br
f&=&-\frac m{2r}+{\mathcal O}\Big(\frac mr\Big)^3\\
g&=&\frac m{2r}-\frac 1{16}\Big(\frac mr\Big)^2+{\mathcal O}\Big(\frac mr\Big)^3.
\er
Hence, up to ${\mathcal O}\Big(\frac mr\Big)^3$, the following relation holds
\begin{equation}\label{3-6}
g=-f-\frac 14 f^2.
\end{equation}
We will seek the time depended solution of the field equation (\ref{3-1}) 
in the diagonal form (\ref{3-3})  with the functions 
$f=f(x^i,t)$ and $g=g(x^i,t)$. \\
Let us assume that the static relation (\ref{3-6}) holds for the 
time dependent metric as well. \\
Thus, in order to determine the metric up to 
${\mathcal O}(\frac mr)^3$, it is enough to find the function $f$. 
This way, by simplifying the assumptions above, the solution of (\ref{3-1}) 
is, approximately, dependent on one function $f$.
If this is needed we can look at  the scalar covariant equation  
\begin{equation}\label{3-7}
R=0.
\end{equation}
Let us first insert the diagonal metric (\ref{3-3}) in (\ref{3-7}). 
The calculations result in 
\begin{equation}\label{3-8}
R=-6e^{-2f}(\ddot {g}-\dot{f}\dot{g}+2\dot{g}^2)+2e^{-2g}\Big(\triangle f+2\triangle g +
(\nabla f)^2+\nabla f\cdot\nabla g+(\nabla g)^2\Big)
\end{equation}
($\dot{f}$ is used for for the time derivative of $f$).\\ 
Insert the relation (\ref{3-6}) to this equation. 
The result is 
\begin{equation}\label{3-9}
R=3e^{-2f}\Big((2-f)\ddot f -(7+5f+f^2)\dot f^2\Big)
+e^{2f+1/2f^2}\Big(-2(1+f)\triangle f+(f+\frac 12 f^2)(\nabla f)^2\Big).
\end{equation}
Estimate the terms of this equation. 
Note, first, that the the term containing 
$\triangle f$ vanishes for the static as well as the time  
dependent solution obtained 
from it by Lorentz type transformations. 
As for the gradient term, it is estimated by 
$$f(\nabla f)^2\sim \frac 1{r^2}{\mathcal O}(\frac mr)^3.$$
(Note that the equation does not include the term 
 $\frac 1{r^2}{\mathcal O}(\frac mr)^2$ which may come 
from the expression $(\nabla f)^2$.)
Thus, up to $\frac 1{r^2}{\mathcal O}(\frac mr)^3$, the curvature 
scalar is
\begin{equation}\label{3-10}
R=6\ddot{f}-2\triangle f.
\end{equation}
Be rescaling the coordinates we obtain the Lorentz invariant equation
\begin{equation}\label{3-11}
R=2 \ \square f=0.
\end{equation}
Recall that the truly dynamical variable are the metric components 
determined by the function $\vp=e^{2f}$. 
The equation for this function is 
\begin{equation}\label{3-12}
\square \vp=\frac 1{\vp}(\nabla \vp)^2.
\end{equation}
Note that the function $\vp=1+\frac mR{\mathcal O}(1)$. 
Hence up to $\frac 1{r^2}{\mathcal O}(\frac mr)^3$ the equation (\ref{3-12}) 
takes the form 
\begin{equation}\label{3-13}
\square \vp=(\nabla \vp)^2.
\end{equation}
This scalar equation is Lorentz invariant and non-linear. 
In the next section it will be shown  that the algorithm is applicable 
for this equation. 
\sect{A Non-linear scalar model}
Let a flat Minkowskian $4D$-space with a metric 
$\eta_{ab}=\eta^{ab}=diag(1,-1,-1,-1)$ and a scalar field $f=f(x)$ be given. 
 Consider, first, the wave equation - a linear second order equation for 
the field $f$ 
\begin{equation}\label{w-eq}
\square f=*d*d f=0.
\end{equation}
It has a unique 1-singular at the origin spherical symmetric 
and asymptotically zero solution 
\begin{equation}\label{w-sol1}
f=\frac mR.
\end{equation}
It is natural to interpret this solution as the field of a 
pointwise body located at the origin. 
The linearity of the field equation (\ref{w-eq}) yields 
the existence of   $N$-singular asymptotically zero solution
\begin{equation}\label{w-sol2}
f=\sum^n_{i=1}\frac {m_i}{|\bR_i|}, \qquad \bR_i=\bbr - \bbr_i.
\end{equation}
The solution (\ref{w-sol2}) have to be interpreted as a field 
produced by a static configuration of bodies  of masses $m_i$ 
located at the points $\bbr_i$. 
This interpretation is, of course,  not physical  because 
it completely ignores the interaction between the particles. 
One may attempt to  apply now an instantaneous Lorentz transformation 
(with a time dependent velocity) of the static solution in order 
to produce a solution with interactions. 
For that, let the singular point move on an arbitrary trajectory $\bpsi(t)$.
Consider the function
\begin{equation}\label{s-sol3}
f=\sum_i \frac {m_i}{|\bR_i|},
\end{equation}
where
\begin{equation}\label{lor2}
\bR_i=(\bbr - \bbr_i)-\a_i\dbpsi_i (\dbpsi_i,(\bbr-\bbr_i))-\b_i \bpsi_i,
\end{equation}
and the Lorentz parameters $\a_i$ and $\b_i$ are functions only 
of $|\dbpsi_i|^2$
\begin{equation}\label{lor3}
\b_i=\frac 1 {\sqrt{1-|\dbpsi_i|^2}}, \qquad\qquad 
\a_i=\frac 1{|\dbpsi_i|^2}\Big(1-\frac 1 {\sqrt{1-|\dbpsi_i|^2}}\Big)
\end{equation}
This is the form of  Lorentz transformation for an arbitrary velocity 
(Cf. Appendix A).\\
For $\bpsi_i=\dbpsi_i t$, at the point $\bR_i=0$  
\begin{equation}\label{2-6}
 \bbr - \bbr_i=\dbpsi_i \Big(\a_i(\dbpsi_i,(\bbr-\bbr_i))+\b_i t\Big).
\end{equation}
So
\begin{equation}\label{2-7}
(\dbpsi_i,(\bbr-\bbr_i))=|\dbpsi_i|^2\Big(\a_i(\dbpsi_i,(\bbr-\bbr_i))+
\b_i t\Big).
\end{equation}
Or
\begin{equation}\label{2-8}
(\dbpsi_i,(\bbr-\bbr_i))=\frac{|\dbpsi_i|^2\b_i t}{1-\a_i |\dbpsi_i|^2}.
\end{equation}
Substituting this relation in (\ref{2-6}) we obtain the equation 
motion of  the singularity as
\begin{equation}\label{2-9}
\bbr-\bbr_i=\frac{\b_i}{1-\a_i |\dbpsi_i|^2}\dbpsi_i t=\dbpsi_i t,
\end{equation}
which is the motion of a free particle.\\
Calculate approximately the d'Alembertian of the function (\ref{s-sol3}),  
in particular, omit the time derivatives of $\a_i$ and $\b_i$ 
 to obtain (Cf. Appendix B)
\begin{equation} 
\square f=\sum_{i=1}^n \frac{m_i\b_i}{R_i^3}(\ddbpsi_i,\bR_i) +
\mathcal O(\ddbpsi\dbpsi).
\end{equation}
Thus, for the field equation (\ref{w-eq})  $f$ is a solution if and only 
if  $\ddbpsi_i=0$.\\
Consequently, the linear equation describes a free inertial motion 
of an arbitrary system of singularities.\\
Consider now  a nonlinear covariant field equation  
\begin{equation} 
d*d \ \!\! f=kd \ \!\! f\wedge *d \ \!\! f,
\end{equation}
or, in coordinate form
\begin{equation} \label{s-eq}
\square f= k\eta^{ab}f_{,a}f_{b}.
\end{equation}
It is easy to see that  the the field equation (\ref{s-eq}) can be 
transformed to the field equation (\ref{w-eq}) by the  redefinition 
of the scalar field 
\begin{equation}\label{trans}
\vp=e^{-kf}.
\end{equation}
If the field $f$ satisfies the equation (\ref{s-eq}) the new field  
$\vp$  satisfies (\ref{w-eq}).  The transformation (\ref{trans}) 
can be used in order to derive a close form solution to the nonlinear 
equation (\ref{s-eq}). 
\begin{equation}\label{s-sol}
f=-\frac 1k \ln{\Big(1+k\frac{m}{|\bR|}\Big)}.
\end{equation}
In order to have a singularity only at the origin we have  to 
require the constants $k$ and $m$ to be positive. \\
The solution (\ref{s-sol}) is singular at the origin and vanishes 
at the infinity. It's long-distance expansion begins with $ m/r$. 
In the limit case $k\to 0 $ the solution (\ref{s-sol}) approaches 
the Newtonian potential $f\to \frac mr$.\\
Because of the  transformation  (\ref{trans}) the equation 
(\ref{s-eq}) has also a static $N$-singular points solution 
\begin{equation}\label{s-sol0}
f=-\frac 1k \ln{\Big(1+k\sum^N_{i=1} \frac{{}^{(i)}m}{{}^{(i)}R}\Big)},
\end{equation}
where ${}^{(i)}\bR=\bbr-\bbr_i$  
and $\bbr_i$ are the radius vectors of the $N$  fixed points.\\
The solution (\ref{s-sol0}) is static. 
Let us look for an approximate dynamic one. 
In order to construct an approximate  solution take 
the superposition of the fields 
\begin{equation}\label{s-sol1}
f=-\frac 1k \sum^N_{i=1}\ln{\Big(1+k \frac{{}^{(i)}m}{{}^{(i)}R}\Big)}.
\end{equation}
Let us apply the  Lorentz transformation with  variable velocities.  
Now the vector $\bR$ c.f. (\ref{lor2}) is time dependent.
Calculate the  linear part of the equation (Cf. Appendix C) to obtain
\begin{equation} 
\square f=\sum^n_{i=1}\Big( \frac {{}^{(i)}m{}^{(i)}\b}{{}^{(i)}R^3}
\cdot\frac{({}^{(i)}\bR,{}^{(i)}\ddbpsi)}{1-k\frac{{}^{(i)}m}{{}^{(i)}R}}
-\frac {{}^{(i)}m^2k}{{}^{(i)}R^4}
\frac 1 {(1-k\frac{{}^{(i)}m}{{}^{(i)}R})^2}\Big).
\end{equation}
By Appendix D the nonlinear part is:
\brn
k\eta^{ab}f_{,a}f_{,b}&=&-k\sum_i\frac {{}^{(i)}m^2k}{{}^{(i)}R^4}\frac 1 {(1-k\frac{{}^{(i)}m}{{}^{(i)}R})^2}
-\sum_{i\ne j}\frac {\frac {{}^{(i)}m}{{}^{(i)}R^3}}{1-k\frac{{}^{(i)}m}{{}^{(i)}R}}\cdot\frac {\frac {{}^{(j)}m}{{}^{(j)}R^3}}{1-k\frac{{}^{(j)}m}{{}^{(j)}R}}\Big[({}^{(i)}\bR,{}^{(j)}\bR)+\\
&&({}^{(i)}\dbpsi,{}^{(j)}\bR)({}^{(i)}\dbpsi,{}^{(i)}\bR) \Big({}^{(i)}\b{}^{(j)}\b+
{}^{(i)}\a+{}^{(j)}\a-{}^{(i)}\a{}^{(j)}\a({}^{(i)}\dbpsi,{}^{(j)}\dbpsi)\Big)\Big].
\ern
The field equation takes the form
\br\label{2-85}
&&\sum^n_{i=1} \frac {{}^{(i)}m{}^{(i)}\b}{{}^{(i)}R^3}\cdot\frac{({}^{(i)}\bR,{}^{(i)}\ddbpsi)}{1-k\frac{{}^{(i)}m}{{}^{(i)}R}}=-k\sum_{i\ne j}\frac {\frac {{}^{(i)}m}{{}^{(i)}R^3}}{1-k\frac{{}^{(i)}m}{{}^{(i)}R}}\cdot\frac {\frac {{}^{(j)}m}{{}^{(j)}R^3}}{1-k\frac{{}^{(j)}m}{{}^{(j)}R}}\Big[({}^{(i)}\bR,{}^{(j)}\bR)+\nonumber\\
&&({}^{(i)}\dbpsi,{}^{(j)}\bR)({}^{(i)}\dbpsi,{}^{(i)}\bR) \Big({}^{(i)}\b{}^{(j)}\b+
{}^{(i)}\a+{}^{(j)}\a-{}^{(i)}\a{}^{(j)}\a({}^{(i)}\dbpsi,{}^{(j)}\dbpsi)\Big)\Big]+{\mathcal O}(\ddot{\psi}\dot{\psi}).
\er
For the approximation of  slow motions it is
\begin{equation}\label{2-86}
\sum^n_{i=1} \frac {{}^{(i)}m}{{}^{(i)}R^3}\cdot\frac{({}^{(i)}\bR,{}^{(i)}\ddbpsi)}{1-k\frac{{}^{(i)}m}{{}^{(i)}R}}=-k\sum_{i\ne j}\frac {\frac {{}^{(i)}m}{{}^{(i)}R^3}}{1-k\frac{{}^{(i)}m}{{}^{(i)}R}}\cdot\frac {\frac {{}^{(j)}m}{{}^{(j)}R^3}}{1-k\frac{{}^{(j)}m}{{}^{(j)}R}}({}^{(i)}\bR,{}^{(j)}\bR)+{\mathcal O}(\ddot{\psi}\dot{\psi})+\frac 1{R^2}{\mathcal O}(\frac mR)^2.
\end{equation}
The two sides of this equation are functions of the variable $x$. 
Consider the $p$-th singularity.
Take an arbitrary point $x$ close to this singularity. 
It follows that
\begin{equation}\label{2-87}
{}^{(p)}\bR\to 0 \text{\quad and\quad } 
{}^{(i)}\bR\to \bar{R}_{ip}\text{\quad for\quad }
i\ne p,
\end{equation}
where $\bar{R}_{ip}$ is a vector from the point $i$ to the point $p$.\\ 
In the LHS of the equation (\ref{2-86}) there is one singular term
\begin{equation}\label{si1}
\frac{{}^{(p)}m}{{}^{(p)}R^3}\frac{({}^{(p)}\bR,{}^{(p)}\ddbpsi)}{1-k\frac{{}^{(p)}m}{{}^{(p)}R}}.
\end{equation}
The singular term in the RHS of (\ref{2-86}) is
\begin{equation}\label{si2}
-k\frac {\frac {{}^{(p)}m}{{}^{(p)}R^3}}{1-k\frac{{}^{(p)}m}{{}^{(p)}R}}\sum_{j\ne p}\frac {\frac {{}^{(j)}m}{{}^{(j)}R^3}}{1-k\frac{{}^{(j)}m}{{}^{(j)}R}}({}^{(p)}\bR,{}^{(j)}\bR).
\end{equation}
The terms (\ref{si1}) and (\ref{si2}) are ${\mathcal O}(R^{-2})$ near 
the singularity. 
When these are inserted to the RHS and LHS of (\ref{2-86}), respectively, 
the remainder will be ${\mathcal O}(R^{-1})$ if only if: 
\begin{equation}\label{2-89}
({}^{(p)}\bR,{}^{(p)}\ddbpsi)=-k\sum_{j\ne p}\frac {\frac {{}^{(j)}m}{{}^{(j)}R^3}}{1-k\frac{{}^{(j)}m}{{}^{(j)}R}}({}^{(p)}\bR,{}^{(j)}\bR)
\end{equation}
This is the only way to diminish the strength of the singularity 
at (\ref{2-86}).\\
Take into account that  the point $x$ is still arbitrary. 
Hence (\ref{2-89}) is valid only if 
\begin{equation}
{}^{(p)}\ddbpsi=-k\sum_{j\ne p}\frac {\frac {{}^{(j)}m}{{}^{(j)}R^3}}
{1-k\frac{{}^{(j)}m}{{}^{(j)}R}}{}^{(j)}\bR
\end{equation}
For the limiting values in (\ref{2-87}) 
\begin{equation}
{}^{(p)}\ddbpsi=-k\sum_{j\ne p}\frac {{}^{(j)}m\bR_{jp}}{R_{jp}^3}\frac 1 {1-k\frac{{}^{(j)}m}{R_{jp}}}
\end{equation}
The second fraction differs from 1, significantly, only 
for small distances comparable to the Schwarzschild radius $r=km$. 
Thus, it can be  neglected. \\
It follows that
 \begin{equation}\label{2-92}
{}^{(p)}\ddbpsi=-k\sum_{j\ne p}\frac {{}^{(j)}m\bR_{jp}}{R_{jp}^3}
\end{equation}
For a system of two singular points  
\begin{equation}\label{2-93}
{}^{(1)}\ddbpsi=-k\frac {{}^{(2)}m\bR_{21}}{R_{21}^3}
\end{equation}
For $k<0$ (\ref{2-92})  and (\ref{2-93}) result in attraction 
between the particles. The absolute value of $k$ is unimportant, 
since it amounts to the rescaling of the mass.\\
This way Newton's law is obtained.
\sect{The Embedding problem}                            
The trajectories obtained are approximate. 
At this point, two avenues are open. 
The first one, which is adopted by EIH is to get higher order approximations 
to the trajectories. 
This procedure is also used in the PPN approach.
By these methods, the successive approximations become highly 
singular near the particle trajectories. \\
The second avenue is to embed the singularities in a field satisfying the 
field equations. 
For that purpose, the successive approximations should add regular terms 
(and, possible, low order singular terms) near the trajectories.  
In this article we just touch upon the method to be developed. 
Consider a first order solution i.e. 
the trajectories are taken to be fixed and the calculations is done only 
for the first order.
For the scalar model the desired field will be a solution 
$\vp$ of (\ref{s-eq}),
\begin{equation}\label{4-1}
\vp=e^{-f+g},
\end{equation}
where $f$ is defined by (\ref{s-sol1}) and $g$ is regular 
at the vicinity of  all the singularities (``multi Green function''). 
Let us present a formal argument that shows that the construction is possible. 
Take, approximately $f={\mathcal O}(\frac m r)$It follows from the computations in the 
previous section that 
\begin{equation}\label{4-3}
(\square f-\eta^{ab}f_{,a}f_{,b})=mF
\end{equation}
(\ref{s-eq}) is satisfied when 
\begin{equation}\label{4-2}
\square g-2\eta^{ab}f_{,a}g_{,b}+\eta^{ab}g_{,a}g_{,b}=-mF.
\end{equation}
Newton's law holds, thus $F={\mathcal O}(\frac 1 r)$. 
Consider for $F$ two examples:
$$ 
F=\frac m r + \text{regular terms}
$$ 
or 
$$ 
F=\frac {m<r,a>^2} {r^3} + \text{regular terms}
$$ 
In the first case, take $g^{(1)}=-\frac 12 cr^2$, in the second case ( a is a constant vector) take 
$g^{(1)}=-\frac 12 c<r,a>r$. In both cases the remainder in the left hand side of 
(\ref{s-eq}) will be regular.
\begin{appendix}
\sect{Lorentz transformations}
The Lorentz transformations are usually written in a very special case when the axes of two reference systems are parallel and one of the systems moves relative to the other with a velocity parallel to the $x$-axis
\footnote{Recall that we use the system of units with $c=1$} 
\begin{equation} \label{lor-1}
x=\a(\tx+v\tt), \quad y=\ty, \quad z=\tz, \quad t=\a(\tt+v\tx),
\end{equation}
with the Lorentz parameter 
\begin{equation} \label{lor-2}
\a=(1-v^2)^{-\frac 12}.
\end{equation}
It is well known that the Lorentz transformations are non-commutative. 
Consequently, a general transformation with an arbitrary directed vector of velocity can not be generated as a successive application of three orthogonal transformations (relative to the axes). \\
In the sequel, a formula for Lorentz transformation for general velocity vector is exhibited. It is hard to believe that such formula does not exists in the literature. The authors, however, could not find a reference for it.\\
Consider a reference system moving with an arbitrary directed velocity $\mathbf{v}$ with the axes parallel to the corresponding axes of a rest reference system. Consider a vector $\bbr$ to an arbitrary point in space. The projection of the vector $\bbr$ on the direction $\bv$ will be 
\begin{equation} \label{proj}
P_{\bv}\bbr=\frac{\bv(\bv,\bbr)}{v^2}.
\end{equation}
Expand the vector $\bbr$ as 
\begin{equation} \label{}
\bbr=P_{\bv}\bbr+N_{\bv}\bbr.
\end{equation}
For a Lorentz transformation
\br
P_{\bv}\tbr&=&\a(P_{\bv}\bbr +\bv t),\\
N_{\bv}\tbr&=&N_{\bv}\bbr. 
\er
Consequently, the transform of the spatial coordinates is
 \begin{equation} \label{}
\tbr=\a(P_{\bv}\bbr +\bv t)+N_{\bv}
\end{equation}
or, explicitly
 \begin{equation} \label{lor1n}
\tbr=\bbr+\bv\Big(\a t-\frac{(1-\a)}{v^2}(\bbr,\bv)\Big).
\end{equation}
The change of the time coordinate is also governed only by the tangential part of the vector $\bbr$
 \begin{equation} \label{}
\tt=\a(t+|P_{\bv}\bbr|v)=\a\Big(t+(P_{\bv}\bbr,\bv)\Big)
\end{equation}
or explicitly
\begin{equation} \label{lor2n}
\tt=\a\Big(t+(\bbr,\bv)\Big).
\end{equation}
In the special case of a motion parallel to the axis $x$ the relations (\ref{lor1n}), (\ref{lor2n})  reduce to the ordinary form of Lorentz transformation (\ref{lor-1}).\\
Therefore an arbitrary  Lorentz transformation takes the form
\begin{equation} \label{a12}
\left\{\begin{array}{ll}
&\tt=\a\Big(t+(\bbr,\bv)\Big)\\
&\tbr=\bbr+\bv\Big(\a t-\frac{(1-\a)}{v^2}(\bbr,\bv)\Big)
\end{array}\right.
\end{equation}
Taking the derivative  of (\ref{a12}) the law of transformations of the differentials is obtained
\begin{equation} \label{}
\left\{\begin{array}{ll}
&d\tt=\a\Big(dt-v_jdx^j\Big)\\
&d\tx^i=dx^i+v^i\Big(\a dt+\frac{(1-\a)}{v^2}v_jdx^j\Big)
\end{array}\right.
\end{equation}
This  can be represented in a matrix form by
 \begin{equation} \label{}
d\tx=Adx \qquad <===> \qquad d\tx^\mu={A^\mu}_\nu dx^\nu
\end{equation}
with
\begin{equation} \label{}
A=
\left(\begin{array}{cccc}
\a &\a v^1 &\a v^2 & \a v^3\\
\a v^1 &1+(1-\a)\frac {v_1^2}{v^2} &(1-\a)\frac {v_1v_2}{v^2} &(1-\a)\frac {v_1v_3}{v^2} \\
\a v^2 &(1-\a)\frac {v_1v_2}{v^2} &1+(1-\a)\frac {v_2^2}{v^2}  &(1-\a)\frac {v_1v_3}{v^2} \\
\a v^3 &(1-\a)\frac {v_1v_3}{v^2} &(1-\a)\frac {v_2v_3}{v^2} &1+(1-\a)\frac {v_3^2}{v^2} 
\end{array}\right)
\end{equation}
For our purposes it is enough to have the second order $(\frac 1{c^2})$ approximation of the transforms (\ref{lor1n}), (\ref{lor2n}). Use
 \begin{equation} \label{}
\a=\frac 1{\sqrt{1-v^2}} = 1+\frac 12 v^2 +O(v^4)
\end{equation}
to obtain
\begin{equation} \label{}
\left\{\begin{array}{ll}
&\tbr=\bbr+\bv t+\bv(\bbr,\bv)\\
&\tt=(1-\frac 12 v^2)t+(\bbr,\bv).
\end{array}\right.
\end{equation}
Consequently the approximation of the transformation law for the 
differentials is 
\begin{equation} \label{}
\left\{\begin{array}{ll}
&d\tt=(1+\frac 12 v^2)dt-v_jdx^j\\
&d\tx^i=dx^i+v^i\Big(dt-\frac 12 v_jdx^j\Big).
\end{array}\right.
\end{equation}
Thus  the matrix of transformation is 
\begin{equation} \label{}
A=
\left(\begin{array}{cccc}
1+\frac 12v^2  &v^1    & v^2                &  v^3\\
 v^1 &1-\frac 12 v_1^2   &-\frac 12 {v_1v_2}  &\frac 12 {v_1v_3} \\
 v^2 &-\frac 12 {v_1v_2} &1-\frac 12 {v_2^2}  &-\frac 12 {v_1v_3} \\
 v^3 &-\frac 12 {v_1v_3}    &\frac 12 {v_2v_3}   &1-\frac 12 {v_3^2}.
\end{array}\right).
\end{equation}
\sect{1-point ansatz}
Calculate the d'Alembertian of the field $f$ of a moving particle  $f$: 
\begin{equation}
f=\frac m{R},
\end{equation}
where
\begin{equation}
\bR=(\bbr - \bbr_0)-\a\dps (\dbpsi,(\bbr-\bbr_0))-\b \p,
\end{equation}
with $\p=\p(t)$ and $\a$ and $\b$ are functions of $|\dbpsi|^2$:
\begin{equation}
\b=\frac 1 {\sqrt{1-|\dbpsi|^2}}, \qquad\qquad \a=\frac 1{|\dbpsi|^2}\Big(1-\frac 1 {\sqrt{1-|\dbpsi|^2}}\Big)
\end{equation}
By straightforward calculation
\br
\bR_t&=&-2\a'\Big((\dbpsi,\ddbpsi)\dbpsi (\dbpsi,\bbr-\bbr_0)\Big)-\a\Big(\ddbpsi(\dbpsi,\bbr-\bbr_0)\Big)\nonumber\\
&&-\a\Big(\dbpsi (\ddbpsi,\bbr-\bbr_0)\Big)
-2\b'\Big((\dbpsi,\ddbpsi) \p\Big) -\b \dbpsi 
\er
and
\br
\bR_{tt}&=&-4\a''\Big((\dbpsi,\ddbpsi)^2 (\dbpsi,(\bbr-\bbr_0))\Big)
-2\a'\Big(|\ddbpsi|^2 (\dbpsi,(\bbr-\bbr_0))\Big)\nonumber\\&&
-2\a'\Big((\dbpsi,\dddbpsi) (\dbpsi,(\bbr-\bbr_0))\Big)
-2\a'\Big((\dbpsi,\ddbpsi) (\ddbpsi,(\bbr-\bbr_0))\Big)\nonumber\\&&
-4\b''\Big((\dbpsi,\ddbpsi)^2 \bpsi \Big)-2\b'\Big(|\ddbpsi|^2 \bpsi  \Big)
-2\b'\Big((\dbpsi,\dddbpsi) \bpsi\Big) -2\b'\Big((\dbpsi,\ddbpsi) \dbpsi\Big) \nonumber\\&&
-2\b'\Big( (\dbpsi,\ddbpsi)\dbpsi \Big)
-\b \ddbpsi. 
\er
Thus
\br
f_t&=&-\frac m{R^3}(\bR_t,\bR)=-\frac m{R^3}\bigg(
-2\a'\Big((\dbpsi,\ddbpsi)(\dbpsi,\bR) (\dbpsi,\bbr-\bbr_0)\Big)-\nonumber\\&&
\a\Big((\ddbpsi,\bR)(\dbpsi,\bbr-\bbr_0)\Big)
-\a\Big((\dbpsi,R) (\ddbpsi,\bbr-\bbr_0)\Big)\nonumber\\
&&-2\b'\Big((\dbpsi,\ddbpsi) (\p ,\bR)\Big)-\b \Big((\dbpsi ,R)\Big)\bigg)
\er
In the first approximation (for velocities that are small  with respect to the speed of the light) we can take
\begin{equation}
\bR_t=-\b\dbpsi, \qquad \bR_{tt}=-\b\ddbpsi.
\end{equation}
Consequently,
\begin{equation}
f_t=\frac m{R^3}\b (\dbpsi ,R)
\end{equation}
In general, the derivatives of $\a$ and $\b$ contribute terms that are quadratic in $\dbpsi$ and its derivatives. Hence, they can  be neglected.
Thus, the second time derivative, to the same accuracy, is 
\br
f_{tt}&=&\frac{m\b}{R^3}\Big((\ddbpsi ,\bR)+(\dbpsi ,\bR_t)\Big)
-3\frac{m\b}{R^5}(\dbpsi ,\bR)(\bR_t,\bR)
\er
Substitute
\begin{equation}
\bR_t=-\b \dbpsi 
\end{equation}
to get
\br
f_{tt}&=&\frac{m\b}{R^3}\Big(
(\ddbpsi ,\bR)-\b \ (\dbpsi ,\dbpsi )\Big)
 \ + \ 3\frac{m\b^2}{R^5}(\dbpsi ,\bR)^2\nonumber\\
&=&\frac{m\b}{R^3}(\ddbpsi ,\bR) \ 
+ \  \ \frac{m\b^2}{R^5}\Big(3(\dbpsi ,\bR)^2-R^2|\dbpsi|^2\Big).
\er
As for the spatial derivatives 
\begin{equation}
\bR_x=\e_1-\a \dot{\psi}_1\dbpsi,  \qquad \qquad \bR_{xx}=0
\end{equation}
\begin{equation}
f_x=-\frac{m}{R^3}(\bR_x,\bR)
\end{equation}
\begin{equation}
f_{xx}=-\frac{m}{R^3}(\bR_x,\bR_x) \ + \ 3\frac{m}{R^5}(\bR_x,\bR)^2
\end{equation}
\begin{equation}
\triangle \ f= 3\frac{m}{R^5}\Big((\bR_x,\bR)^2+(\bR_y,\bR)^2+(\bR_z,\bR)^2\Big)-
\frac{m}{R^3}\Big(R_x^2+R_y^2+R_z^2\Big)
\end{equation}
Since
\begin{equation}
(\bR_x,\bR)=R_1-\a\dot{\psi}_1(\dbpsi,R)
\end{equation}
we get
\begin{equation} 
(\bR_x,\bR)^2+(\bR_y,\bR)^2+(\bR_z,\bR)^2=R^2-2\a(\dbpsi,R)^2+\a^2|\dbpsi|^2(\dbpsi,R)^2
\end{equation}
and
\begin{equation} 
R_x^2+R_y^2+R_z^2=3-2\a |\dbpsi|^2+\a^2|\dbpsi|^4
\end{equation}
Thus the Laplacian is 
\begin{equation} 
\triangle f=\frac{m}{R^5}(\a^2|\dbpsi|^2-2\a)\Big(3(\dbpsi,R)^2-|\dbpsi|^2R^2\Big)
\end{equation}
And, for the d'Alembertian
\begin{equation} 
\square f=\frac{m\b}{R^3}(\ddbpsi ,\bR)+\frac{m}{R^5}(\b^2+2\a-\a^2|\dbpsi|^2)\Big(3(\dbpsi ,\bR)^2-R^2|\dbpsi|^2\Big)
\end{equation}
Using the expressions for the functions $\a,\b$ we obtain, approximately, 
\begin{equation} 
\square f=\frac{m\b}{R^3}(\ddbpsi ,\bR)
\end{equation}
\sect {The leading part}
From the calculations above
\begin{equation} 
\bR_t=-\b\dbpsi, \qquad \bR_{tt}=-\b\ddbpsi.
\end{equation}
Thus 
\begin{equation} 
f_t=-\frac m{R^3}\frac{(\bR,\bR_t)}{1-k\frac{m}{R}}=
\frac {m\b}{R^3}\frac{(\bR,\dbpsi)}{1-k\frac{m}{R}}.
\end{equation}
And
\br
f_{tt}=\frac{m\b}{R^5}\cdot\frac{3\b(\bR,\dbpsi)^2+R^2(\bR,\ddbpsi)-\b R^2|\dbpsi|^2}
{1-k\frac{m}{R}}+
\frac{m^2\b^2 k}{R^6}\cdot\frac{(\bR,\dbpsi)^2}{(1-k\frac{m}{R})^2}.
\er
As for the spatial derivatives we have
\begin{equation}
\bR_x=\e_1-\a\dbpsi(\dbpsi,\e_1), \qquad \bR_{xx}=0,
\end{equation}
where $e_1$ is a unit vector on the $x$ axis.
\begin{equation}
f_x=-\frac m{R^3}\cdot\frac{(\bR,\bR_x)}{1-k\frac{m}{R}}
\end{equation}
\begin{equation}
f_{xx}=3\frac m{R^5}\cdot\frac{(\bR,\bR_x)^2}{1-k\frac{m}{R}}-
\frac m{R^3}\cdot\frac{(\bR_x,\bR_x)}{1-k\frac{m}{R}}+
\frac {km^2}{R^6}\cdot\frac{(\bR,\bR_x)^2}{(1-k\frac{m}{R})^2}.
\end{equation}
Thus 
\br
\triangle f&=&3\frac m{R^5}\cdot\frac{(\bR,\bR_x)^2+(\bR,\bR_y)^2+(\bR,\bR_z)^2}
{1-k\frac{m}{R}}-\nonumber\\
&&\frac m{R^3}\cdot\frac{(\bR_x,\bR_x)+(\bR_y,\bR_y)+(\bR_z,\bR_z)}{1-k\frac{m}{R}}+\nonumber\\
&&\frac {km^2}{R^6}\cdot\frac{(\bR,\bR_x)^2+(\bR,\bR_y)^2+(\bR,\bR_z)^2}{(1-k\frac{m}{R})^2}
\er
Substitute the value of $\bR_x$ to get
\br
\triangle f&=&3\frac m{R^5}\cdot\frac {R^2-2\a(\dbpsi,\bR)^2+\a^2|\dbpsi|^2(\dbpsi,\bR)^2}{1-k\frac{m}{R}}-
\frac m{R^5}\cdot\frac{3-2\a|\dbpsi|^2 +\a^2|\dbpsi|^4}{1-k\frac{m}{R}}+\nonumber\\&&\frac {km^2}{R^6}\cdot\frac{R^2-2\a(\dbpsi,\bR)^2+\a^2|\dbpsi|^2(\dbpsi,\bR)^2}{(1-k\frac{m}{R})^2}
\er
Thus the second order l.h.s. of the equation is
\br
\square f&=&\frac{m\b}{R^5}\cdot\frac{3\b(\bR,\dbpsi)^2+R^2(\bR,\ddbpsi)-\b R^2|\dbpsi|^2}{1-k\frac{m}{R}}+
\frac{m^2\b^2 k}{R^6}\cdot\frac{(\bR,\dbpsi)^2}{(1-k\frac{m}{R})^2}\nonumber\\&&
-3\frac m{R^5}\cdot\frac {R^2-2\a(\dbpsi,\bR)^2+\a^2|\dbpsi|^2(\dbpsi,\bR)^2}{1-k\frac{m}{R}}+\frac m{R^3}\cdot\frac{3-2\a|\dbpsi|^2 +\a^2|\dbpsi|^4}{1-k\frac{m}{R}}-
\nonumber\\&&\frac {km^2}{R^6}\cdot\frac{R^2-2\a(\dbpsi,\bR)^2+\a^2|\dbpsi|^2(\dbpsi,\bR)^2}{(1-k\frac{m}{R})^2}\nonumber\\
&=&\frac {m\b}{R^3}\cdot\frac{(\bR,\ddbpsi)}{1-k\frac{m}{R}}+
\frac m{R^5}(\b^2+2\a-\a^2|\dbpsi|^2)\cdot\frac{(3(\bR,\dbpsi)^2-R^2|\dbpsi|^2)}{1-k\frac{m}{R}}+\nonumber\\&&
\frac {m^2k}{R^6}\cdot\frac{-R^2+(\b^2+2\a-\a^2|\dbpsi|^2)(\bR,\dbpsi)^2}{(1-k\frac{m}{R})^2}
\er
Using the relation $\b^2=\a^2|\dbpsi|^2-2\a$ we obtain
\begin{equation} 
\square f=\frac {m\b}{R^3}\cdot\frac{(\bR,\ddbpsi)}{1-k\frac{m}{R}}-\frac {m^2k}{R^4}\frac 1 {(1-k\frac{m}{R})^2}
\end{equation}
As for the quadratic r.h.s. 
\br
\eta^{ab}f_{,a}f_{,b}&=&\frac {m^2\b^2}{R^6}\cdot\frac{(\bR,\dbpsi)^2}{(1-k\frac{m}{R})^2}-
\frac {m^2}{R^6}\cdot\frac{(\bR,\bR_x)^2+(\bR,\bR_y)^2+(\bR,\bR_z)^2}{(1-k\frac{m}{R})^2}\nonumber\\
&=&\frac {m^2\b^2}{R^6}\cdot\frac{(\bR,\dbpsi)^2}{(1-k\frac{m}{R})^2}-
\frac {m^2}{R^6}\cdot\frac{R^2-2\a(\dbpsi,\bR)^2+\a^2|\dbpsi|^2(\dbpsi,\bR)^2}{(1-k\frac{m}{R})^2}\nonumber\\
&=&-\frac {m^2}{R^4}\frac 1 {(1-k\frac{m}{R})^2}+\frac {m^2}{R^6}(\b^2+2\a-\a^2|\dbpsi|^2)\frac{(\bR,\dbpsi)^2}{(1-k\frac{m}{R})^2}
\er
Thus, to the first order,
\begin{equation} 
\eta^{ab}f_{,a}f_{,b}=-\frac {m^2}{R^4}\frac 1 {(1-k\frac{m}{R})^2}
\end{equation}
\sect {The quadratic part}
Calculate the nonlinear part
\begin{equation} 
f_t=-\sum^n_{i=1}\Big(\frac {({}^{(i)}\bR_t,{}^{(i)}\bR)}{1-k\frac{{}^{(i)}m}{{}^{(i)}R}}\cdot \frac {{}^{(i)}m}{{}^{(i)}R^3}\Big)=
\sum^n_{i=1}\Big(\frac {({}^{(i)}\dbpsi,{}^{(i)}\bR)}{1-k\frac{{}^{(i)}m}{{}^{(i)}R}}\cdot \frac {{}^{(i)}m{}^{(i)}\b}{{}^{(i)}R^3}\Big).
\end{equation}
Thus
\begin{equation} 
(f_t)^2=\sum^n_{i,j=1}\frac {\frac {{}^{(i)}m{}^{(i)}\b}{{}^{(i)}R^3}}{1-k\frac{{}^{(i)}m}{{}^{(i)}R}}\cdot\frac {\frac {{}^{(j)}m{}^{(j)}\b}{{}^{(j)}R^3}}{1-k\frac{{}^{(j)}m}{{}^{(j)}R}}({}^{(i)}\dbpsi,{}^{(i)}\bR)({}^{(j)}\dbpsi,{}^{(j)}\bR).
\end{equation}
As for the spatial derivatives 
\begin{equation} 
f_x=-\sum^n_{i=1}\frac {({}^{(i)}\bR_x,{}^{(i)}\bR)}{1-k\frac{{}^{(i)}m}{{}^{(i)}R}}\cdot 
\frac {{}^{(i)}m}{{}^{(i)}R^3}
\end{equation}
\br \label{2-79}
(\nabla f,\nabla f)&=&\sum^n_{i,j=1}\frac {\frac {{}^{(i)}m}{{}^{(i)}R^3}}{1-k\frac{{}^{(i)}m}{{}^{(i)}R}}\cdot\frac {\frac {{}^{(j)}m}{{}^{(j)}R^3}}{1-k\frac{{}^{(j)}m}{{}^{(j)}R}}\Big(({}^{(i)}\bR_x,{}^{(i)}\bR)({}^{(j)}\bR_x,{}^{(j)}\bR)+\nonumber\\
&&({}^{(i)}\bR_y,{}^{(i)}\bR)({}^{(j)}\bR_y,{}^{(j)}\bR)+({}^{(i)}\bR_z,{}^{(i)}\bR)({}^{(j)}\bR_z,{}^{(j)}\bR)\Big)
\er
Using the relation
\begin{equation} 
{}^{(i)}\bR_x=\e_1-{}^{(i)}\a{}^{(i)}\dbpsi({}^{(i)}\dbpsi,\e_1)
\end{equation}
and writing 
\begin{equation} 
({}^{(i)}\bR_x,{}^{(i)}\bR)=(\e_1,{}^{(i)}\bR)-{}^{(i)}\a({}^{(i)}\dbpsi,{}^{(i)}\bR)({}^{(i)}\dbpsi,\e_1)
\end{equation}
we obtain
\br
&&\Big(({}^{(i)}\bR_x,{}^{(i)}\bR)({}^{(j)}\bR_x,{}^{(j)}\bR)\Big)=
\Big((\e_1,{}^{(i)}\bR)(\e_1,{}^{(j)}\bR)\Big)-\nonumber\\
&&\qquad\qquad\quad{}^{(j)}\a\Big((\e_1,{}^{(i)}\bR)({}^{(j)}\dbpsi,{}^{(j)}\bR)({}^{(j)}\dbpsi,\e_1)\Big)-\nonumber\\
&&\qquad\quad\qquad
{}^{(i)}\a\Big((\e_1,{}^{(j)}\bR)({}^{(i)}\dbpsi,{}^{(i)}\bR)({}^{(i)}\dbpsi,\e_1)\Big)+
\nonumber\\
&&\qquad\quad\qquad{}^{(i)}\a{}^{(j)}\a({}^{(i)}\Big(\dbpsi,{}^{(i)}\bR)({}^{(i)}\dbpsi,\e_1)({}^{(j)}\dbpsi,{}^{(j)}\bR)({}^{(j)}\dbpsi,\e_1)\Big).
\er
Thus the brackets in (\ref{2-79}) are
\br
\Big({\quad }\Big)&=&({}^{(i)}\bR,{}^{(j)}\bR)-{}^{(j)}\a\Big(({}^{(j)}\dbpsi,{}^{(i)}\bR)({}^{(j)}\dbpsi,{}^{(j)}\bR)\Big)-\nonumber\\
&&{}^{(i)}\a\Big(({}^{(i)}\dbpsi,{}^{(j)}\bR)({}^{(i)}\dbpsi,{}^{(i)}\bR)\Big)+\nonumber\\
&&{}^{(i)}\a{}^{(j)}\a\Big(({}^{(i)}\dbpsi,{}^{(i)}\bR)({}^{(i)}\dbpsi,{}^{(j)}\dbpsi)({}^{(j)}\dbpsi,{}^{(j)}\bR)\Big).
\er
Consequently, the r.h.s. of the field equation is
\brn
k\eta^{ab}f_{,a}f_{,b}&=&-\sum^n_{i,j=1}\frac {\frac {{}^{(i)}m}{{}^{(i)}R^3}}{1-k\frac{{}^{(i)}m}{{}^{(i)}R}}\cdot\frac {\frac {{}^{(j)}m}{{}^{(j)}R^3}}{1-k\frac{{}^{(j)}m}{{}^{(j)}R}}\Big[({}^{(i)}\bR,{}^{(j)}\bR)+\\
&&({}^{(i)}\dbpsi,{}^{(j)}\bR)({}^{(i)}\dbpsi,{}^{(i)}\bR) \Big({}^{(i)}\b{}^{(j)}\b+
{}^{(i)}\a+{}^{(j)}\a-{}^{(i)}\a{}^{(j)}\a({}^{(i)}\dbpsi,{}^{(j)}\dbpsi)\Big)\Big]
\ern
\end{appendix}


\end{document}